%% file: main.tex
\pdfoutput=1

\documentclass[letterpaper,twocolumn,10pt]{article}
\usepackage{usenix2019_v3,epsfig}
\usepackage[utf8]{inputenc}
\usepackage{textcomp}
\usepackage{authblk}
\usepackage{enumitem}
\usepackage{multirow}
\setlist{nosep} % Reduce itemize/enumerate spaces
\setlist[itemize,1]{leftmargin=\dimexpr 1.4em}
\setlist[enumerate,1]{leftmargin=\dimexpr 1.4em}

\date{}

\title{\Large \bf NNStreamer: Stream Processing Paradigm for Neural Networks,\\Toward Efficient Development and Execution of On-Device AI Applications}

\author[1]{\rm MyungJoo~Ham}  % Writing (Most) + Code (HIGH) + PI
\author[1]{\rm Ji~Joong~Moon} % Writing (Skeleton, Ch. 5.1) + Code (HIGH)
\author[1]{\rm Geunsik~Lim}   % Writing (Skeleton, Chapter 2, Bib) + Code (Low)
\author[1]{\rm Wook~Song}     % Writing (Ch. 5.2) + Experiment + Code (Low)
\author[1]{\rm Jaeyun~Jung}   % Code (HIGH)
\author[1]{\rm Hyoungjoo~Ahn} % Code (Mid)
\author[1]{\rm Sangjung~Woo}  % Experiment + Code (Low)
\author[1]{\rm Youngchul Cho} % Experiment (ARS) / From Project:NS
\author[2]{\rm Jinhyuck~Park} % Code (Low)
\author[1]{\rm Sewon~Oh}      % Code (Low)
\author[1,3]{\rm Hong-Seok~Kim}  % Discussion / Writing (Low)

\affil[1]{Samsung Research, Samsung Electronics}
\affil[2]{Biotech Academy, Samsung BioLogics}
\affil[3]{Left the affiliation}
\affil[1,2]{\{myungjoo.ham, jijoong.moon, geunsik.lim, wook16.song, jy1210.jung, hello.ahn, sangjung.woo, rams.cho, jinhyuck83.park, sewon.oh, hongse.kim\}@samsung.com}
\begin{document}
\maketitle

% Use the following at camera-ready time to suppress page numbers.
% Comment it out when you first submit the paper for review.
\thispagestyle{empty}

\begin{abstract}
\input{sections/abstract.tex}
\end{abstract}

\section{Introduction}\label{S_Intro}
\input{sections/S01_Intro.tex}

\section{Related Work}\label{S_RWork}
\input{sections/S02_RelatedWork.tex}

\section{Approaches}\label{S_Approach}
\input{sections/S03_Approach.tex}

\section{Implementation}\label{S_Implementation}
\input{sections/S04_Implementation.tex}

\section{Evaluations}\label{S_Evaluation}
\input{sections/S05_Evaluation.tex}

\section{Future Work}\label{S_FutureWork}
\input{sections/S06_FutureWork.tex}

\section{Conclusion}
\input{sections/S07_Conclusions.tex}

%%%%%%%%%%%%%%%%%%%%%%%%%%%%%%%%%%%%%%%%%%%%%%%%%%%%%%%%% 
%comment/@leemgs: it's trivial. I am not sure that we have to write an open source contribution here without the "Conclusion" section.
% @mj: this is per the template of ATC. We can merge this into conclusion if
% we do not have enouch spaces.

%% Hide some info for blind reviews. Make more info available for CAM-READY version.
\section*{Availability}
\textit{Nnstreamer} is open source software
developed in a public GitHub repository.
% More information about \textit{nnstreamer} and example applications are available in GitHub.
%We release binary packages for Tizen and Android developers.
Ubuntu 16.04 and 18.04 users may use PPA to install \textit{nnstreamer} and keep it updated:

\begin{verbatim}
$ sudo add-apt-repository ppa:nnstreamer/ppa
$ sudo apt-get update
$ sudo apt-get install nnstreamer
\end{verbatim}

%%% Acknowledgement will be added for CAM-READY version, not in the for-review version.
%\section*{Acknowledgements}
% MTCNN-ROS & Project:NS Developers
% Sponsors (SR managers)
% GStreamer community

% Page limit is 11 pages including footnotes.
% but not including the bib only.

{\normalsize \bibliographystyle{acm}
\bibliography{nnstreamer}}

\end{document}

%% file: sections/abstract.tex
We propose \textit{nnstreamer}, a software system
that handles neural networks as filters of stream pipelines,
applying the stream processing paradigm to neural network applications.
A new trend with the wide-spread of deep neural network applications is on-device AI;
i.e., processing neural networks directly on mobile devices or edge/IoT devices instead of cloud servers.
Emerging privacy issues, data transmission costs, and operational costs signifies the need for on-device AI especially when a huge number of devices with real-time data processing are deployed.
\textit{Nnstreamer} efficiently handles neural networks with complex data stream pipelines on devices, improving the overall performance significantly with minimal efforts.
Besides, \textit{nnstreamer} simplifies the neural network pipeline implementations and allows reusing off-shelf multimedia stream filters directly; thus it reduces the developmental costs significantly.
\textit{Nnstreamer} is already being deployed with a product releasing soon and is open source software applicable to a wide range of hardware architectures and software platforms.

%Many usage scenarios of on-device AI address real-time data streams from sensors of the device itself: e.g., cameras and microphones.
%For such data streams, users and governments do not want manufacturers to access the live feeds of the streams, manufactures and operators do not want to send and receive such huge traffic of streams incurring high latency, and they also do not want to spend excessively for transmissions and servers especially when they have extremely many devices deployed.

% \subsection{Proposal 1}
% 1. Deep learning이 퍼져가고 있음
% 근데, On-Device AI가 중요해짐
% On-Device AI 중에서도 stream 처리 형태가 매우 중요할 것임
% ??? 실제 현실에 쓴다고 보면 - 여러 NN들이 사실 복잡한 pattern으로...

% 2. NN을 stream filter로 접근하도록 해주는 Stream Processing FW을 제안하겠음.

% 서로 다른 NN FW을 지원, mix & match 가능
% 복잡한 pipeline 구성 가능
% 성능

% 3. 근데, 우리 FW은 GStreamer를 활용. 그래서 X, Y, Z 추가 장점이 있음
% 그런데 NN 지원 관점에서 필요한 기능들을 plugin 형태로 구현, XXX 했음
% 이미 널리 쓰이고 있음 - 안정성, 개발자 풀
% 다양한 media type들, pre/post procesing 지원

%% file: sections/S01_Intro.tex
% 1. Intro of on-device intelligence. What is on-device intelligence? Why it is important and emerging?

We have witnessed the wide spread of deep neural networks and their applications in the last decade.
With ever growing computing power of embedded or edge devices,
neural networks are being adopted to such devices, further assisted by AI accelerators~\cite{APPLE_A12X, chen2018tvm, Moreau:vta-tvm, ionica2015movidius, Jouppi:TPU, npu-isvlsi, ignatov2018ai, exynos9-9820}. 
In mobile phone industry, this trend has already become obvious enough to be adopted by major manufacturers including Samsung and Apple~\cite{exynos9-9820, APPLE_A12X}.
Running intelligence mechanisms such as deep neural networks directly in edge devices including consumer electronics is often called as on-device AI~\cite{MITTechReview}.

On-device AI is becoming more attractive to manufacturers because of the following three advantages:
\begin{enumerate}
\item Avoid data privacy and protection issues by keeping the data in user devices without uploading to cloud servers.
\item Reduce data transmission latency of high-bandwidth real-time stream data including live video feeds from cameras.
\item Reduce operating cost by off-loading computation to devices from servers. This cost is often neglected; it is, however, significant if billions of devices are to be deployed.
\end{enumerate}
We do not discuss the potential advantage, distributing workloads across edge devices, because it is not in the scope of this paper, but of future work.

% 2. The characteristics and issues of on-device intelligence
%% a. Online stream data. High data bandwidth.
%% b. (Often) Limited computing power (handheld / home appliances / TV / ...)
%% c. Latency and throughput may matter more (live video manipulation / ...?)

On-device AI achieves such advantages by processing directly in the nodes where data exist so that data transmission is reduced and sensitive data are kept inside.
However, on-device AI induces significant challenges.
Normally, edge devices including mobile phones, TVs, home appliances, and IoT devices have limited computing power compared to workstations of conventional AI systems.
In the meanwhile, on-device AI applications often require short response time or high processing rates; e.g., AR Emoji~\cite{SamsungAREmoji}, Animoji~\cite{AppleAnimoji}, and offline speech recognition and translations.
Besides, on-device AI applications often use sensors incurring data stream with high bandwidth: video cameras.

% 2-1. Secondary (derived) issues from these issues or uncovered new emerging issues (could be more important)
%% a. (need to make conjectures) Complex pipelines are to be constructed along with multiple neural networks in a device or an applications. And constructing, maintaining, and dynamically reconfiguring such pipelines is a real PITA.
%% b. Synchornization

On-device AI applications are becoming more complex, incurring even more challenges.
A neural network may process multiple input streams and
multiple neural networks may process streams simultaneously.
Besides, networks may share input data, and an input of a network may be outputs of other networks.
For example, inputs of a facial recognition and an emotion recognition may share outputs of an object recognition, saving computing resources.
Composing systems with multiple networks allows to train smaller networks (more efficient) and to save training costs.
However, stream pipelines for such systems may become highly complicated with interconnections of networks and sensors along with fluctuating latency and synchronization issues.
Another challenge is that the topology of pipelines may be dynamic.

We have observed another issue in practice; developers may use their own neural network frameworks such as TensorFlow and Caffe or their own tweaked version of such frameworks.
With stable releases of applications, we may enforce developers to share a framework or to provide independent binaries without the need for any neural network frameworks.
However, for researching and prototyping, developers want to experiment with various frameworks while testers still need to execute the whole system integrated.
Thus, we need to compose such complicated interactions between neural networks and sensors along with multiple neural network frameworks.

% 3. Possible mitigation / approaches for the issues
%% a. Hardware acceleration. ==> we still nee
%% Nah.. let's skip this and see how it looks.

% 4. Our approach, the "concept" or "principles"

We apply the stream processing paradigm~\cite{stephens1997survey} to on-device AI systems.
Our approach is to handle a neural network as a stream filter and then, to construct the system as a stream pipeline.
We propose data standards for tensors and containers of tensors, which allow varying neural network frameworks in a pipeline along with conventional stream filters, conventional media types (video, audio, and text), and tensors.
In order to construct complicated pipelines, we also propose various stream path manipulators with synchronization policies and tensor operators.

% 5. Brief description of what/how we've implemented + 6. What's the benefit of our implementation & approach

Handling stream pipelines efficiently along with various synchronization methods, filters, and data types has been a main topic of multimedia frameworks for decades.
Among multimedia frameworks, we choose GStreamer~\cite{GStreamer} as the basis because GStreamer offers highly modular architecture, is fully open source software, and is compatible with various operating systems and hardware architectures.
We can also reuse what GStreamer offers; hundreds of media processing plugins and developmental infrastructure.
GStreamer has been widely deployed to desktop computing, video conferencing, broadcasting, and consumer electronics, even including televisions of Samsung and LG, where the multimedia performance is extremely critical.

% 7. How it's evaluated, commercialized, deployed, ...
% They requests to change fall detection to activity recognition (micro activity recognition). (any suggestions?? I don't think "micro" is necessary.)
An experimental product deployed to hospitals and households, which will be commercially released soon, is based on \textit{nnstreamer} to process recognize events from multiple sensors.
Product developers have significantly improved the performance and simplified the code, which have enabled to add more sensors and neural networks with inexpensive hardware and minimal developmental costs.
\textit{Nnstreamer} is open source software licensed with LGPL~\cite{LGPL}.
We plan to use \textit{nnstreamer} for more products in the affiliation including robots and conventional consumer electronics.

%% file: sections/S02_RelatedWork.tex
% On-Device Intelligence and Stream-Inference

\subsection{Multimedia Stream Processing in Practice}

Multimedia stream frameworks have recently evolved to process
high quality audio and video with mobile phones and TVs whose computing resources are limited.

%\begin{itemize}
%\item GStreamer as third party media stream common support for Android/iOS (Pexip) as well.
%\item GStream: A general-purpose data streaming framework on GPU clusters
%\end{itemize}

\textbf{GStreamer}~\cite{GStreamer} is the standard multimedia pipeline framework for Tizen and many Linux distributions.
GStreamer provides APIs and utilities to construct stream pipelines for multimedia applications of various platforms including Linux, Android, Windows, iOS, and macOS.
%Pexip~\cite{pexip-gstpriqueue,pexip-media-arch} has successfully deployed GStreamer-based enterprise applications enabling collaboration between incompatible video and audio technologies.
GStreamer is highly modular; every filter and path control is implemented as a plugin attached in run-time.
GStreamer has been applied to various systems where multimedia performance matters.
BBC uses GStreamer for their broadcasting system\cite{BBCGSTCONF}.
Samsung and LG use GStreamer as the multimedia engine of televisions.
Centricular~\cite{centricular-gstplayer,centricular-sync-multi-room,centricular-new-gststream-api,centricular-opengl-pipeline,centricular-zerocopy} releases GStreamer for TVs, set-top boxes, medical devices, in-vehicle infotainment, video-on-demand, and on-demand streaming solutions.

Another popular framework for Linux, \textbf{FFmpeg}~\cite{FFMPEG} is not modular, but everything is built-in and has limited cross-platform support.
Note that GStreamer can embed FFmpeg as yet another plugin.

% StageFright (Android)
% \begin{itemize}
% \item A comparative study of android and iOS for accessing internet streaming services
% \item The Implementation of Multimedia Decoder Framework for Android on PAC Duo Plateform
% \end{itemize}
% 

% \cite{pacdsp}\cite{comparative-study} are removed from here because these articles are not "introducing StageFright".
\textbf{StageFright}~\cite{StageFright} is the multimedia stream framework of Android.
StageFright depends on Android services and, unlike GStreamer, is not flexible enough for other generic Linux distributions.
Besides, it does not allow to construct arbitrary stream pipelines.

%Yao Liu\cite{comparative-study} analyzed and compared the performance when Android and iOS devices are accessing Internet streaming services. 
%Their analysis provide some insights for the current Android and iOS users, streaming service providers, and mobile mediaplayer developers. 
%On Android devices, a mobile user can access video streaming services from either the mobile browser or applications.
%Starting from Android 2.3 (Gingerbread), a new media player framework called Stagefright is used in Android.
%Similar to AppleCoreMedia, Stagefright also supports Pseudo Streaming by using HTTP for requesting video data. 
% Stagefright would only use a fixed amount of memory despite different video file sizes, and that only a fixed amount of video data would be kept in the buffer.

\textbf{AVFoundation}~\cite{AVFoundation} is the multimedia stream framework of iOS and macOS.
Along with Core ML, a machine learning framework, we may have AVFoundation as an input to Core ML; however, it is not supposed to write arbitrary pipelines with Core ML entities as stream filters.

\textbf{DirectShow}~\cite{directshow-new-media-ach} is the multimedia framework of Windows.
We cannot alter and use DirectShow or AVFoundation for the given purpose because it is proprietary software and supports proprietary platforms only.
%Microsoft's DirectShow~\cite{directshow-new-media-ach} and Windows Driver Model provide the infrastructure for today’s post-production applications and hardware to truly become interoperable.
%This paper describes the architecture, supporting technologies, and their application in post-production scenarios.
%The Windows driver model (WDM) is a new driver model that makes development of low-latency, cross platform drivers more practical.

\subsection{Stream Processing in General}

% Big data stream processing (server/cloud)
% \begin{itemize}
% \item Data streaming algorithms for efficient and accurate estimation of flow size distribution
% \item Streamcloud: An elastic and scalable data streaming system
% \item A client-side statistical prediction scheme for energy aware multimedia data streaming
% \item Fragmental proxy caching for streaming multimedia objects
% \end{itemize}
% 

StreamCloud~\cite{streamcloud} has addressed the bottleneck of a single input node in a stream pipeline.
StreamCloud provides a scalable and elastic stream framework to process large data input and execute pipeline elements in parallel transparently by splitting queries to allocate multiple nodes.
%Note that input sensors of on-device AI usually exist as independent nodes avoiding a single input node bottleneck and GStreamer provides transparent parallelism for pipeline elements as well.

There are studies to help manage large data streams efficiently; however, they do not provide a general stream processing framework for neural networks and complex topology.
Kumar~\cite{Kumar-data-streaming-algo} has proposed a network data stream analysis tool with higher accuracy and lower overhead than prior work.
%@wook predicts -> estimates
In order to reduce power consumption of edge devices receiving multimedia data streams, a client-side statistical prediction scheme \cite{statistical-prediction} estimates the data patterns of incoming data streams and tries to make processors sleep longer. 
A fragmental proxy-caching scheme~\cite{fragmental-proxy-caching} tries to improve the quality of multimedia streams by adjusting caching units with finer granularity.
Apache Flink~\cite{apache-flink} proposes a platform with a universal dataflow engine designed to perform both stream and batch analytics for processing streaming and batch data.

Applying streaming framework for GPUs fits well with general architectures of GPUs.
GStream~\cite{gstream} provides a general-purpose and scalable data streaming framework for GPUs.
%It is inspired by a lack of streaming abstraction dedicated to massively parallel architectures and their suitability to express data parallelism.
%GStream’s strength is in its ease of use and its applicability to a variety of domains not constrained to traditional streaming problems.
Nvidia Tesla architecture \cite{nvidia-tesla} provides a flexible, programmable graphics and high-performance computing along with the Compute Unified Device Architecture (CUDA) platform.
Jiangiang Dong~\cite{birch} demonstrates how modern GPUs and CUDA can improve the performance of BIRCH, one of well-known clustering techniques for streaming data.

\subsection{Stream Processing for Neural Networks}

Nvidia DeepStream~\cite{GSTCONF18_DEEPSTREAM, nvidia-deepstream} provides neural network filters for GStreamer.
DeepStream requires the input and output of a neural network to be conventional media stream.
Other critical issues include: a) DeepStream is proprietary, b) it supports Nvidia hardware only, and c) general neural network frameworks cannot be applied directly.

In GStreamer Conference 2018, other approaches to apply neural networks to GStreamer are introduced as well.
Intel~\cite{GSTCONF18_Intel} proposes an approach similar with DeepStream, sharing the same issues.
Pexip~\cite{GSTCONF18_Pexip} proposes to use OpenCV to apply neural networks as prior work \cite{directshow-opencv}.
This approach is the least invasive method; however, this cannot be applied to neural networks with arbitrary input and output streams or with general neural network frameworks.
RidgeRun~\cite{GSTCONF18_RidgeRun} has proposed an approach somewhat similar with \textit{nnstreamer}.
However, it does not support neural networks with arbitrary input and output streams or path controls and, according to the opened code~\cite{RidgeRunGithub}, it is far from releases.

% The streaming API that is built on top of Flink’s streaming dataflow engine provides the means to keep recoverable state and to partition, transform, and aggregate data stream windows.
% While batch computations are, in theory, a special case of a streaming computations, Flink treats them specially, by optimizing their execution using a query optimizer and by implementing blocking operators that gracefully spill to disk in the absence of memory.
% Under the same unifying framework, a novel online (adaptive) algorithm is developed to obtain multi-way decompositions of low-rank tensors with missing entries and perform imputation as a byproduct

In contrast, we recognize ``tensor'' as a stream data type, not limiting streams to conventional media types (audio, video, and text), allowing outputs of neural networks to be inputs of other networks in general and stream path controls discussed in Chapter~\ref{S_Implementation}.
We provide efficient and high-performing methods to process neural network pipelines in parallel with arbitrary frameworks, hardware, and platforms.
%, but also offers a visualization tool to identify performance bottlenecks in pipelines and zero memory copy operations for the performance.
% Let's not brag too much with the GST tools here.

% End of subsection

%% file: sections/S03_Approach.tex
\begin{figure*}[t]
\begin{center}
\includegraphics[width=0.99\textwidth]{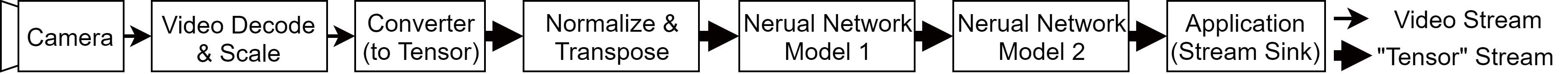}
\end{center}
\caption{Basic linear pipeline with two neural networks and single input and output without path manipulations}
\label{FIG_EXPPLN_Basic_FullLinear}
\end{figure*}

\begin{figure*}[ht]
\centering\includegraphics[width=0.90\textwidth]{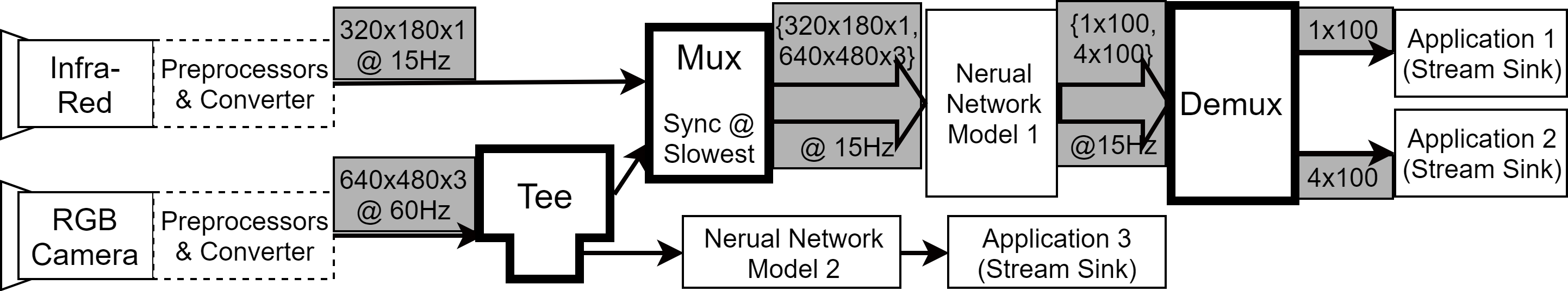}
\caption{Pipeline with isodimensional stream path controls: Tee, Mux, and Demux}
\label{FIG_EXPPLN_Path_KeepTensorDim}
\end{figure*}

We describe the approaches of \textit{nnstreamer} with example pipelines showing the need for each component.
Later in Chapter \ref{S_Implementation}, we show how such components are implemented.

\subsection{Basic Components}\label{SS_BasicComponents}

Neural network frameworks (NNFWs), such as Caffe and TensorFlow, allow developers to describe, train, and execute neural network models.
NNFWs also provide execution environments along with hardware acceleration.

An \textbf{NNFW filter}, ``Neural Network'' in Figure~\ref{FIG_EXPPLN_Basic_FullLinear}, allows to attach a neural network model and its NNFW as a stream filter of a pipeline.
Each NNFW filter may use its own NNFW.
A \textbf{stream sink} transmits the results to external entities such as applications and visualization services.

We need a standard stream data format representing inputs and outputs of neural networks, called ``tensor''.
With the standard, we may run different NNFWs in a single pipeline and implement common data and path manipulators, directing outputs of networks as inputs of others.
Then, we need \textbf{Converters} converting media streams to ``tensor'' streams.

We need \textbf{preprocessors} for tensors and media streams.
A neural network input may have strict requirements; e.g., AlexNet~\cite{2012AlexNet} requires a normalized tensor of 3 x 224 x 224.
To meet such requirements, we need various preprocessors including
conventional media filters (e.g., Video Decode \& Scale in Figure~\ref{FIG_EXPPLN_Basic_FullLinear}) and \textbf{tensor operators} (e.g., Normalize \& Transpose in Figure~\ref{FIG_EXPPLN_Basic_FullLinear}.)

\subsection{Isodimensional Stream Path Control}\label{SS_IdoDim_SPC}
% mux, demux, tee, repo

In Figure~\ref{FIG_EXPPLN_Path_KeepTensorDim}, we show three dimension-preserving stream path controls.
Rectangles with dotted lines represent sets of filters.
Shaded rectangles show data formats.
Thick arrows are streams where each frame has multiple instances of ``tensor''s.

\textbf{Tee} allows multiple paths to share same data.

\textbf{Mux} multiplexes streams into a single stream having multiple instances of ``tensor''s in each frame, referred as ``tensor\textbf{s}''.
In the figure, properties of ``tensors'' are denoted with curly brackets.
Because Mux has a single output stream, its input streams are synchronized by Mux,
which provides multiple synchronization policies.
For example, ``slowest'' synchronizes with the slowest input, ``base'' synchronizes with a specific input, and ``fastest'' generates an output whenever an input is available.
With ``base'' or ``fastest'', Mux may need to reuse input frames from a slower stream.
For example, in Figure~\ref{FIG_EXPPLN_Path_KeepTensorDim}, if the policy is ``fastest'', previous frames from Infra-Red may be reused to meet 60 Hz.

Mux needs to look at time-stamps of incoming stream, stamped by sensors.
Mux needs to choose each frame from the input streams with the closest time-stamp.
For example, if frames of time-stamps, $\{ 14, 30, 49 \}$, are available from Infra-Red and $\{ 29 \}$ from RGB Camera has just arrived, Mux chooses $30$ unless a user specifies otherwise.

\textbf{Demux} demultiplexes an incoming ``tensor\textbf{s}'' stream into multiple ``tensor'' streams as shown in the figure.
We do not need synchronization mechanisms for Demux.

\begin{figure}[t]
\begin{center}
\includegraphics[width=0.95\columnwidth]{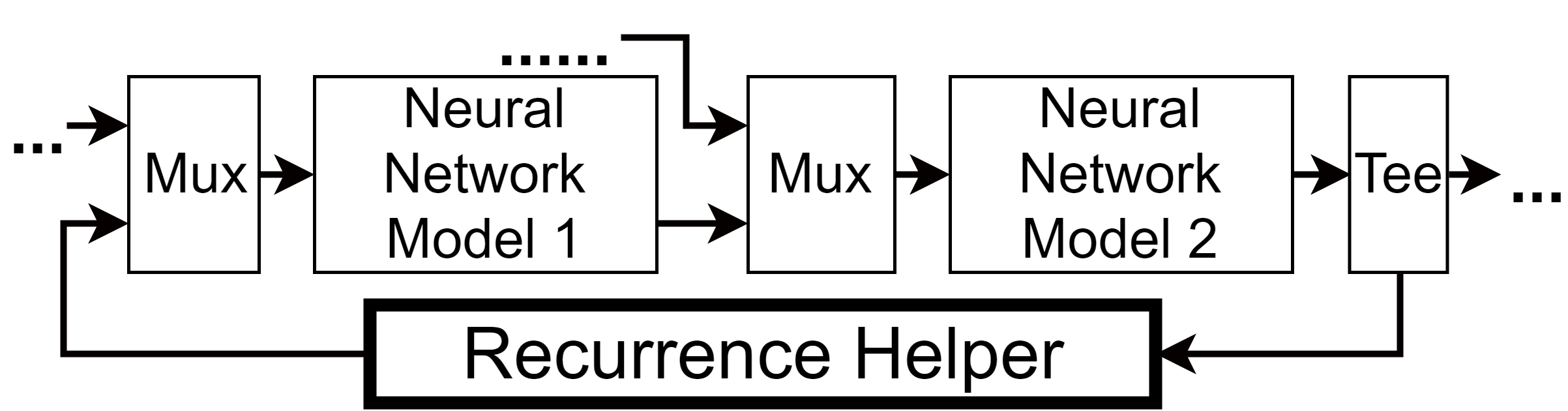}
\end{center}
\caption{Pipeline with external recurrences}
\label{FIG_EXPPLN_Path_RNN}
\end{figure}

Recurrent neural networks (RNN) are popular for handwriting recognition and speech recognition applications.
If a recurrence occurs in a single element of a pipeline without affecting other elements, referred as an \textit{internal recurrence}, it is not exposed to its stream pipeline.
However, as shown in Figure~\ref{FIG_EXPPLN_Path_RNN}, if the recurrence affects and is visible to the pipeline, referred as an \textit{external recurrence}, we need \textbf{Recurrence Helper}.
Recurrent Helper resolves the bootstrapping issue; the output of Model 2 in the figure is not available at the start, which, in turn, makes the input of Model 1 not available and blocks the the whole pipeline.
Besides, if we have meta data streams flowing back and forth as in GStreamer pipelines for QoS, cyclic paths can be prohibited; Recurrence Helper should resolve such conflicts as well.

\subsection{Non-isodimensional Stream Path Control}\label{SS_NONISODIM_SPC}
% merge, split, aggregator

\begin{figure}[t]
\begin{center}
\includegraphics[width=0.98\columnwidth]{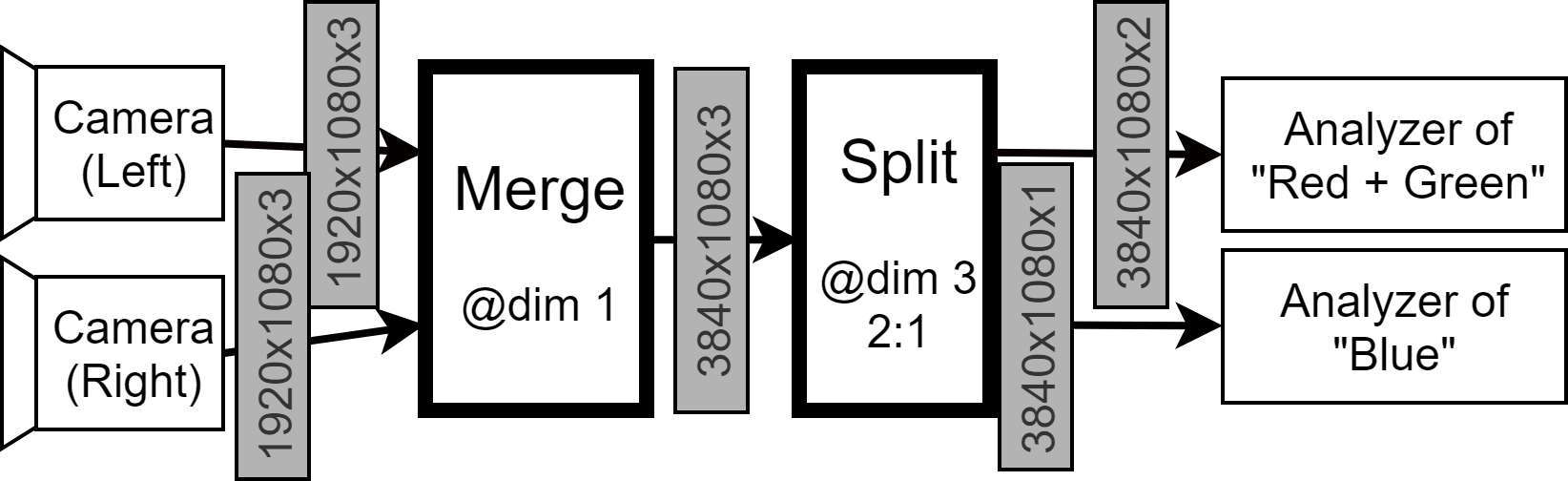}
\end{center}
\caption{Pipeline with Merge and Split}
\label{FIG_EXPPLN_Path_NonIso1}
\end{figure}

Non-isodimensional stream path controls may alter the dimensions of tensors.
Figure~\ref{FIG_EXPPLN_Path_NonIso1} shows two of such controls, \textbf{Merge} and \textbf{Split}.
As shown in the figure, the dimension being merged or splited needs to be specified.
Merge needs synchronization and time-stamp mechanisms like Mux.

% LSTM
\begin{figure}[t]
\begin{center}
\includegraphics[width=0.98\columnwidth]{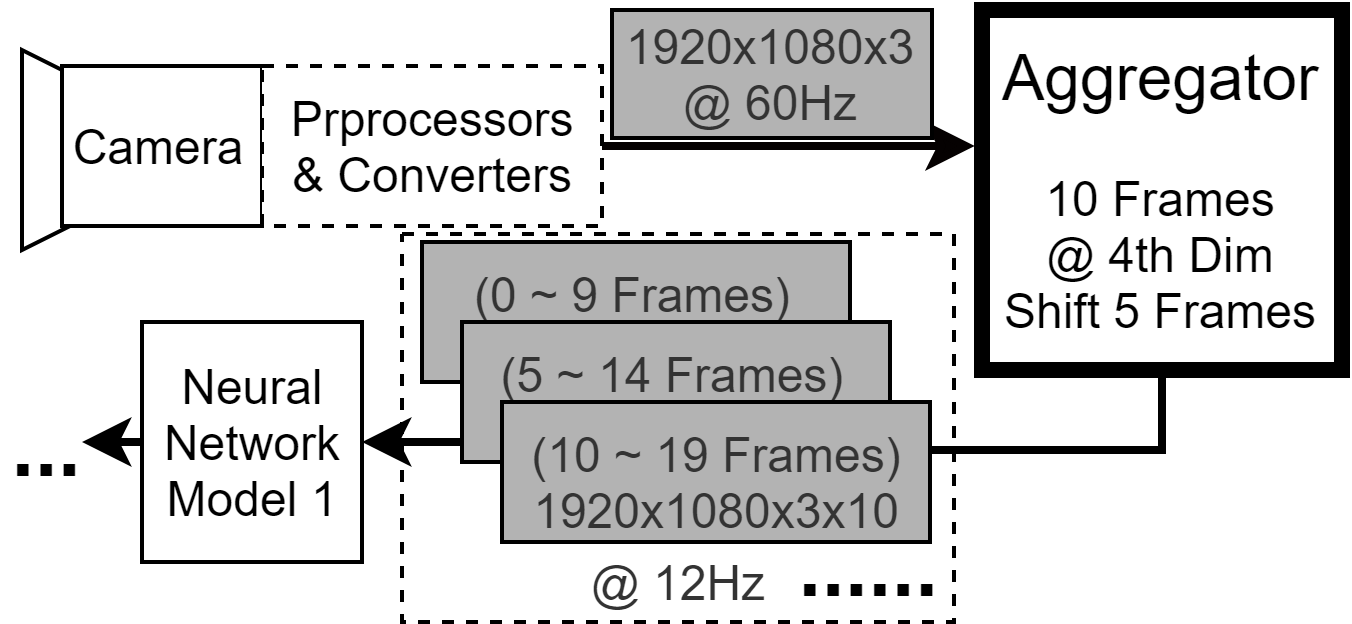}
\end{center}
\caption{LSTM pipeline with Aggregator}
\label{FIG_EXPPLN_Path_Aggregator}
\end{figure}

The long short-term memory (LSTM) is popular for object tracking, machine translation, image captioning, and various applications.
In order to generate inputs for LSTM, we need to aggregate multiple frames of a stream into a single frame: \textbf{Aggregator} in Figure~\ref{FIG_EXPPLN_Path_Aggregator}.
Aggregator merges frames temporally while Mux and Merge merges frames spatially.
Figure~\ref{FIG_EXPPLN_Path_Aggregator} shows an LSTM model processing 10 recent frames simultaneously.
Similar configuration can be seen in activity recognition sensors~\cite{2017NIPS_ActivityDetect}, which use \textit{nnstreamer}, in Chapter~\ref{S_Evaluation_ACTIVITY}.

\subsection{Other Requirements}\label{SS_OtherRequirements}

\begin{figure}[t]
\begin{center}
\includegraphics[width=0.98\columnwidth]{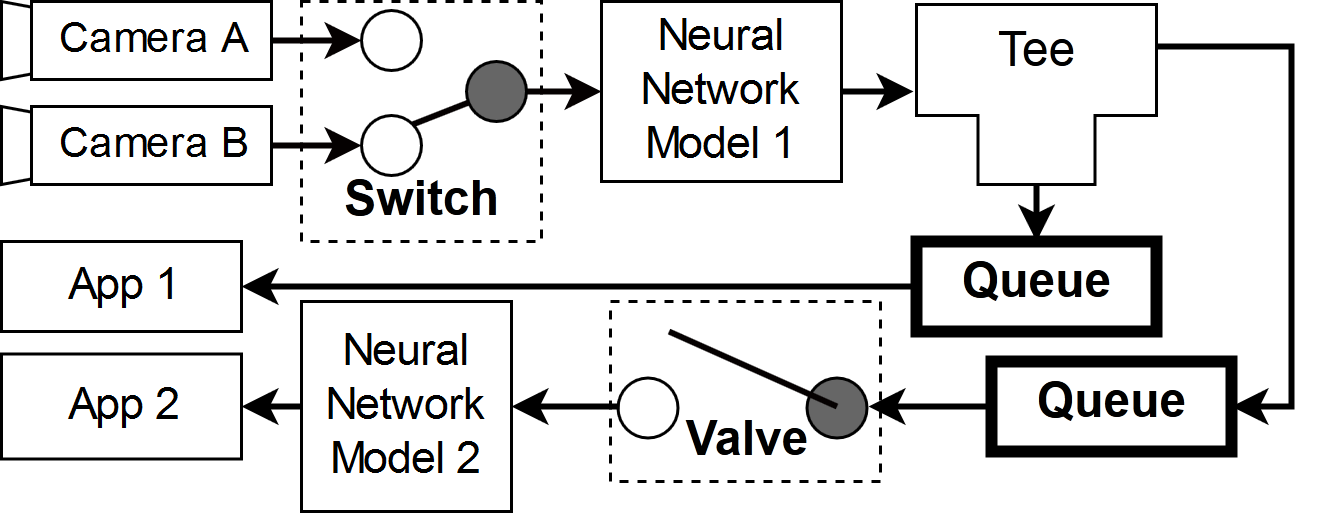}
\end{center}
\caption{Pipeline with queues, a switch, and a valve}
\label{FIG_EXPPLN_MISC}
\end{figure}

From the beginning, we have been cooperating with potential product developers, who have requested the followings:
\begin{itemize}
    \item Dynamic flow control to enable and disable neural networks quickly: \textbf{Valve} in Figure~\ref{FIG_EXPPLN_MISC}.
    \item Change stream sources dynamically and easily in case of sensor faults or mode changes: \textbf{Switch} in Figure~\ref{FIG_EXPPLN_MISC}.
    \item Memcpy-less data transmission between stream filters.
    \item Transparent parallelism. In Figure~\ref{FIG_EXPPLN_MISC}, developers may want to execute \texttt{Model 1}, \texttt{Model 2}, and \texttt{Application 1} in parallel without explicitly implementing threads or synchronization methods.
    \item Dynamic stream pipeline topology. Add, replace, realign, or remove elements of a pipeline.
    \item Manage time-stamps from sensors. With Mux, Merge, and Aggregator, this is not trivial; which time-stamp is valid if we have multiple values for a frame?
\end{itemize}

%% file: sections/S04_Implementation.tex
\textit{Nnstreamer} is not an experimental prototype,
but software with imminent schedule and pressure of commercialization;
thus, it should be ready for tiring quality assurance processes.

Some of authors have been working for the autonomous vehicle project,
where the need for applying stream processing paradigm has emerged.
The vehicle has a lot of heterogeneous sensors and neural networks inter-connected with complex topology, transmitting large data streams, and processed simultaneously.
The huge data transmission overhead between nodes had been the core cause requiring \textit{nnstreamer}.
Then, managing complex and ever-changing data paths, allowing different NNFWs, and parallelism have been incurring the need for \textit{nnstreamer}.

Spinning off this project from the autonomous vehicle project, we have discovered that such needs are common: robots, manufacturing plants, and consumer electronics.
As a result, \textit{nnstreamer} has already had a lot of potential applications in the affiliation from the beginning, which enforces to release partial but practical solutions quickly.

\begin{quote}
    Don't reinvent the wheel, just realign it.\\
    -- Anthony J. D'Angelo
\end{quote}

If we implement yet another stream framework, it would take too much time and effort.
Fortunately, a stable, widely-deployed, and open-source stream framework with rich features already exists: GStreamer.
GStreamer is a multimedia stream framework battle-tested with a lot of services
including BBC~\cite{BBCGSTCONF} and consumer electronics including televisions (both Samsung and LG), where multimedia performance is critical.
Releasing large number of products, a lot of vendors have made sure GStreamer is reliable, high-performing, and rich-featured.
Tizen, which runs a wide range of devices including most types of consumer electronics~\cite{ham2019tizen}, also has GStreamer as its standard multimedia engine.
As a result, most of requirements in Chapter~\ref{SS_OtherRequirements} and the above are already satisfied by GStreamer.

We only need to ``realign'' GStreamer for neural networks and their applications.
We implement what GStreamer is missing: the features described in Chapter~\ref{S_Approach} except for ``Tee'', ``Valve'', ``Switch'', ``Queue'', and conventional media filters.

\subsection{Standard Stream Data Types}

We have designed two new data types for GStreamer, ``{\tt other/tensor}'' and ``{\tt other/tensor}\textbf{s}'', representing a stream of a multidimensional array and a stream of a container of multiple instances of such arrays, respectively.
The definitions of both stream data types in the following paragraphs show how each frame instance of a stream appears.

\begin{verbatim}
other/tensor
    framerate = (fraction) [0/1, 2147483647/1]
    dimension = Dim
    type = Type
  
other/tensors
    num_tensors = [1, 16]
    framerate = (fraction) [0/1, 2147483647/1]
    dimensions = Dims
    types = Types
# num_tensors == # types == # dimensions.

Types = Type | Type,Types
Type = { uint8, int8, uint16, int16, uint32,
       int32, uint64, int64, float32, float64}
Dims = Dim | Dim,Dims
Dim = [1,65535]:[1,65535]:[1,65535]:[1,65535]
# [min, max] denotes the allowed value ranges.
\end{verbatim}

Unlike other approaches~\cite{GSTCONF18_DEEPSTREAM,GSTCONF18_Intel,GSTCONF18_Pexip,GSTCONF18_RidgeRun}, neural networks in \textit{nnstreamer} do not use conventional media types, but use the standard defined above because of the advantages mentioned in Chapter~\ref{SS_BasicComponents}.
For demonstrations, we may assume that the neural network output is a video stream; however, if an application uses the output, it usually wants the raw data, not the rasterized video frames.
More critically, if there is another neural network using the output as its input, conventional media types do not fit often; e.g., an array of label probabilities is neither audio or video.
Besides, a neural network may require inputs preprocessed in the form not compatible with conventional media types or the input itself might be not compatible with conventional media types.
Therefore, standard tensor data stream types are critically required.

\subsection{Stream Filters}

\textit{Nnstreamer}, a GStreamer plugin, v0.1.0-1rc1, has the following elements released as stream filters.

\begin{itemize}
    \item \textbf{tensor\_converter} converts audio, video, text, or arbitrary binary streams to {\tt other/tensor} streams.
    \item \textbf{tensor\_decoder} converts {\tt other/tensor(s)} to video or text stream with assigned sub-plugins.
    \item \textbf{tensor\_filter} invokes a neural network model with the given model path and NNFW name.
    \item \textbf{tensor\_transform} applies various operators to tensors including typecast, add, mul, transpose, and normalize. For faster processing, it supports SIMD instructions and multiple operators in a single filter.
    \item \textbf{tensor\_mux}, \textbf{tensor\_demux}, \textbf{tensor\_merge}, \textbf{tensor\_split}, and \textbf{tensor\_aggregator} support tensor stream path controls.
    \item \textbf{tensor\_reposink} and \textbf{tensor\_reposrc} implement Recurrent Helper of Figure~\ref{FIG_EXPPLN_Path_RNN}. In a pipeline of the figure, tensor\_reposink is attached to ``Tee'' and tensor\_reposrc is attached to the second ``Mux'', cutting the cycle in the pipeline.
    There is a shared repository for the two elements to transmit tensors without GStreamaer stream paths.
\end{itemize}

\begin{figure}[t]
\begin{center}
\includegraphics[width=0.98\columnwidth]{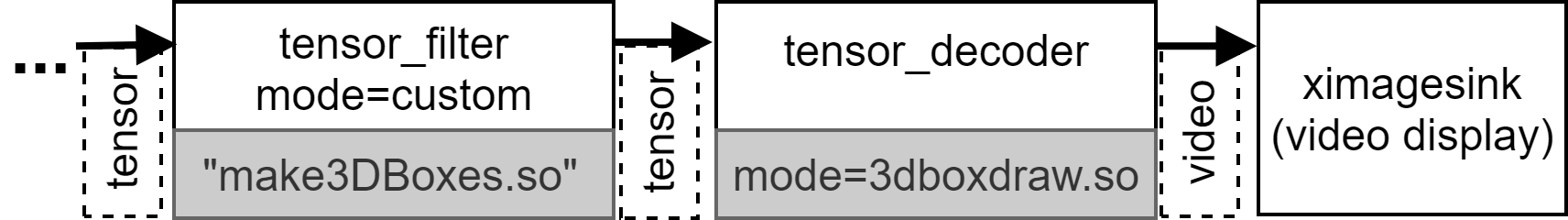}
\end{center}
\caption{User sub-plugins (shaded) applied to elements}
\label{FIG_IMPL_SUBPLUGIN}
\end{figure}

There are two elements allowing adding user created features in run-time: tensor\_filter and tensor\_decoder.
Figure~\ref{FIG_IMPL_SUBPLUGIN} shows how user created neural networks and decoders can be attached to tensor\_filter and tensor\_decoder in a pipeline.
In the figure, ``{\tt make3DBoxes.so}'' is a neural network without NNFWs and ``{\tt 3dboxdraw.so}'' is a user decoder sub-plugin, decoding results of ``{\tt make3DBoxes.so}'' to video streams.
If the tensor\_filter element has ``{\tt mode=tensorflow-lite}'' and ``{\tt model=x.tflite}'', TensorFlow-lite processes ``{\tt x.tflite}''.
\textit{Nnstreamer} provides APIs to implement sub-plugins.
And sub-plugins can be implemented and attached in run-time except for adding a new NNFW to tensor\_filter,
which is to be supported with later releases.

\subsection{Developmental Environments}

A CI system~\cite{TAOS-CI} builds and tests every pull-request in GitHub for Tizen/\{x64,arm64,arm32,x86\} and Ubuntu/x64.
We check \textit{nnstreamer} in Ubuntu/\{arm32,arm64\} daily.
We occasionally check Yocto and Android releases as well.

GStreamer supports various programming languages, hardware architectures, and operating systems.
As a plugin of GStreamer, we ensure the compatibility by writing C89 code and using only the libraries GStreamer uses except for detachable sub-plugins and test cases.
APIs are available for C, C++, .NET, Java, Python, Rust, Perl, Qt, Haskell, D, Guile, Ruby, and Vala.
It supports Linux, Android, Windows, Mac OS X, iOS, BSD, Unix, Solaris, and Symbian along with x86, ARM, MIPS, SPARC and PowerPC.

% ######## Instruction manual
% https://github.sec.samsung.net/STAR/nnstreamer/tree/master/tools/profiling
%
% ######## Example script
% $ gst-launch-1.0 textoverlay name=overlay font-desc="Sans, 26" ! \
%    videoconvert ! ximagesink name=img_test \
%    v4l2src name=cam_src ! videoscale ! video/x-raw,width=640,height=480,format=RGB ! \
%    tee name=t_raw \
%    t_raw. ! queue ! overlay.video_sink \
%    t_raw. ! queue ! videoscale ! video/x-raw,width=224,height=224 ! \
%    tensor_converter ! \
%    tensor_filter framework=tensorflow-lite \
%    model=tflite_model/mobilenet_v1_1.0_224_quant.tflite ! \
%    tensor_decoder mode=image_labeling \
%    option1=tflite_model/labels.txt ! \
%    overlay.text_sink
% 
% ######## Output
% $ gst-report-1.0  pipeline0.gsttrace
% ELEMENT           %CPU   %TIME   TIME
% tensorfilter0      88.4   98.5    57.8 s
% videoscale0         0.8    0.9    524 ms
% overlay             0.2    0.3    154 ms
% videoscale1         0.1    0.1   67.3 ms
% videoconvert0       0.1    0.1   64.8 ms
% img_test            0.0    0.0   16.4 ms
% t_raw               0.0    0.0   16.1 ms
% queue0              0.0    0.0   10.6 ms
% capsfilter0         0.0    0.0   5.52 ms
% tensordec0          0.0    0.0   4.18 ms
% queue1              0.0    0.0   2.66 ms
% tensorconverter0    0.0    0.0   1.72 ms
% capsfilter1         0.0    0.0   1.10 ms
% cam_src             0.0    0.0    370 us
% pipeline0           0.0    0.0      0 ns
% 
% ######### Original source
% \\10.113.111.238\swcsait\papers\tool-tracing-data-figure
%

\begin{figure}[t]
\begin{center}
\includegraphics[width=0.95\columnwidth]{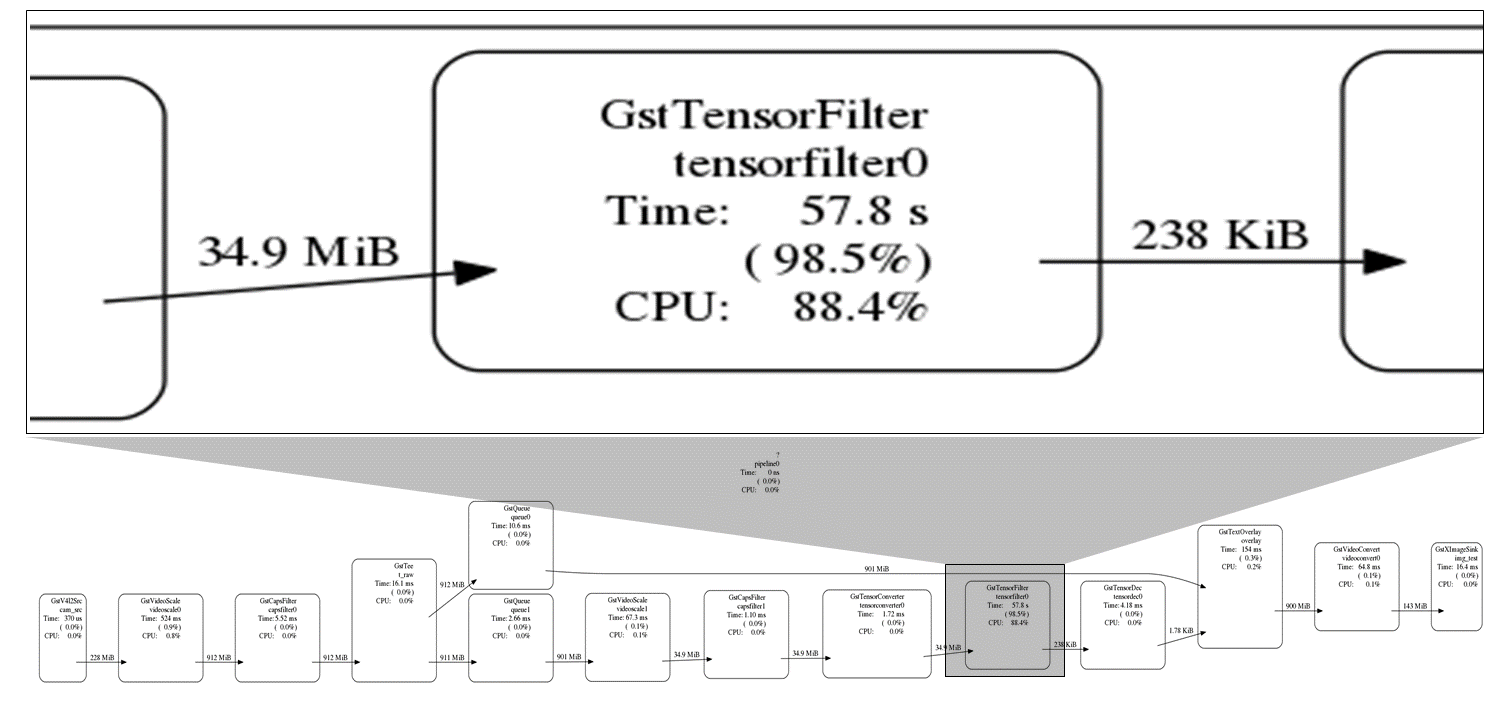}
\end{center}
\caption{Screenshot of profiling an \textit{nnstreamer} pipeline.}
\label{FIG_IMPL_TRACE}
\end{figure}

We can reuse a wide range of stream analysis tools including profilers, tracers, test suites, and debuggers from GStreamer.
Figure~\ref{FIG_IMPL_TRACE} is a screenshot of a tool analyzing a \textit{nnstreamer} pipeline.
We heavily depend on such tools to optimize or to fix bugs of commercialization projects.

%% file: sections/S05_Evaluation.tex
% Experimental Environment
\begin{table*}[tb]
    \centering
    \begin{tabular}{l c c c c}
         & \texttt{A} / Artik 530s & \texttt{B} / Odroid-XU4 & \texttt{C} / 8890 & \texttt{D} / PC\\
         &&&&\vspace{-1em}\\
         \hline
         &&&&\vspace{-1em}\\
    SoC  & Nexell S5P4418 & Samsung Exynos 5422 & Samsung Exynos 8890 & Intel i7-7700 \\
         &&&&\vspace{-1em}\\
    ISA  & armv7 (32 bits) & armv7 (32 bits) & armv8 (64 bits) & x64 (64 bits) \\
         &&&&\vspace{-1em}\\
    CPU Microarchitecture & 4 CA9 & 4 CA15 + 4 CA7 & 4 Mongoose + 4 CA53 & 4 Kaby-Lake\\
         &&&&\vspace{-1em}\\
    CPU Clock Speed & 1.2 GHz & 2 GHz / 1.5GHz & 2.3 GHz / 1.6 GHz & 3.6 GHz \\
         &&&&\vspace{-1em}\\
    DRAM Capacity & 1~GiB DDR3 & 2~GiB LPDDR3 & 4~GiB LPDDR4 & 16~GiB DDR4\\
         &&&&\vspace{-1em}\\
    DRAM Throughput%
    \footnotemark
    & {0.51 / 0.75} & {2.62 / 4.08} &  {5.86 / 6.40} & {18.54 / 13.78}\\
         &&&&\vspace{-1em}\\
    OS \& Linux Kernel & Ubuntu 16.04 / 4.4 & Tizen 5.0 / 4.1 & Tizen 5.0 / 3.18 & Ubuntu 16.04 / 4.15 \\
         \hline
    \end{tabular}
    \caption{Specification of experimented devices. CA denotes Cortex-A. Linux kernels are supplied by the vendors.}
    \label{TBL_SPEC_BOARDS}
\end{table*}

We experiment with \textit{nnstreamer} 0.1.0-1rc1 in devices described in Table~\ref{TBL_SPEC_BOARDS}
for two applications, an activity recognition sensor (ARS) based on \cite{2017NIPS_ActivityDetect} and
Multi-Task Cascaded Convolutional Networks (MTCNN)~\cite{mtcnn-dai2016instance,mtcnn-zhang2016joint}.
Device \texttt{A}, Artik 530s, is the computing engine of ARS consisting of a dynamic vision sensor (DVS)~\cite{2014DVS} and an ultra-wideband (UWB)~\cite{1990UWB} sensor.
Device \texttt{B}, Odroid-XU4, is an inexpensive development board with the SoC of Samsung Galaxy S5.
Device \texttt{A} and \texttt{B} represent mid to low-end embedded devices.
Device \texttt{C} is a development board with Exynos 8890 used by Samsung Galaxy S7 and automotive industry~\cite{8890Automotive}.
This represents high-end embedded devices. 
Device \texttt{D} is a desktop PC.
We do not use GPUs in the experiments.

For ARS, we use Device \texttt{A} and compare an \textit{nnstreamer} pipeline against a conventional implementation as the control.
For MTCNN, we use Device \texttt{B}, \texttt{C}, and \texttt{D} and compare an \textit{nnstreamer} pipeline against ROS~\cite{quigley2009ros}, popular middleware for robotics projects, implementation as the control.
The control represents the product development before \textit{nnstreamer}.

For both applications, \textit{nnstreamer} improves the performance significantly.
The implementation is significantly simplified along with the improved code quality as well.

\subsection{Experiments with ARS}
\label{S_Evaluation_ACTIVITY}

\begin{figure*}[ht]
\centering\includegraphics[width=0.95\textwidth]{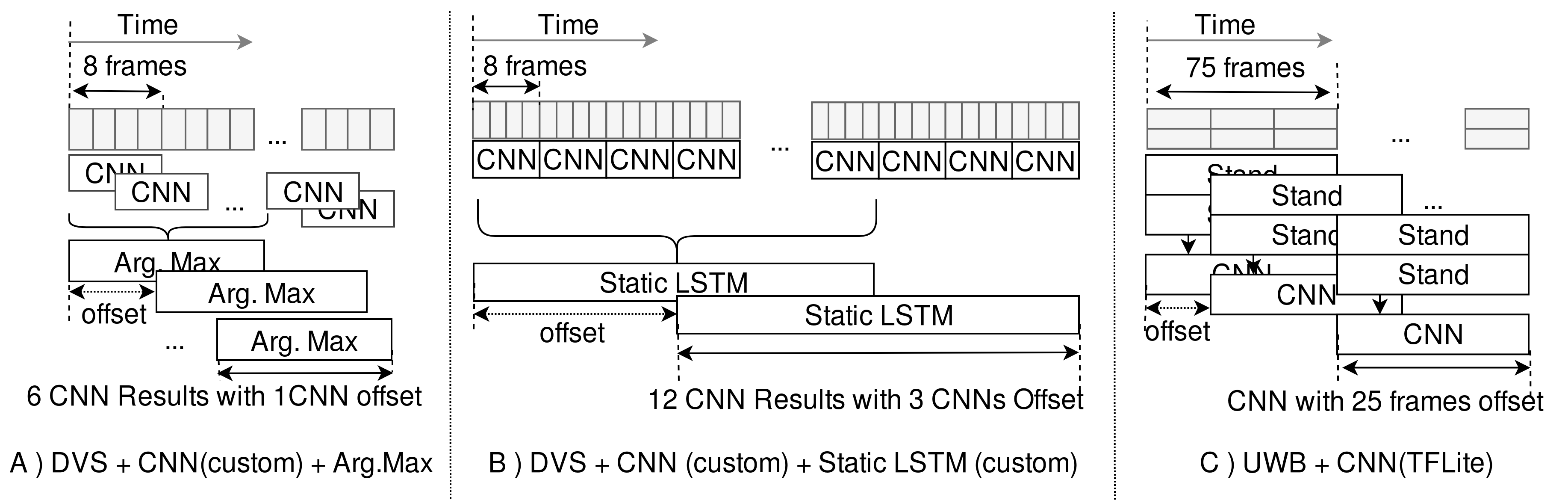}
\caption{Neural network algorithms of activity recognition sensors (ARS)}
\label{FIG_EVAL_Activity_Detection}
\end{figure*}

\begin{figure*}[ht]
\centering\includegraphics[width=0.95\textwidth]{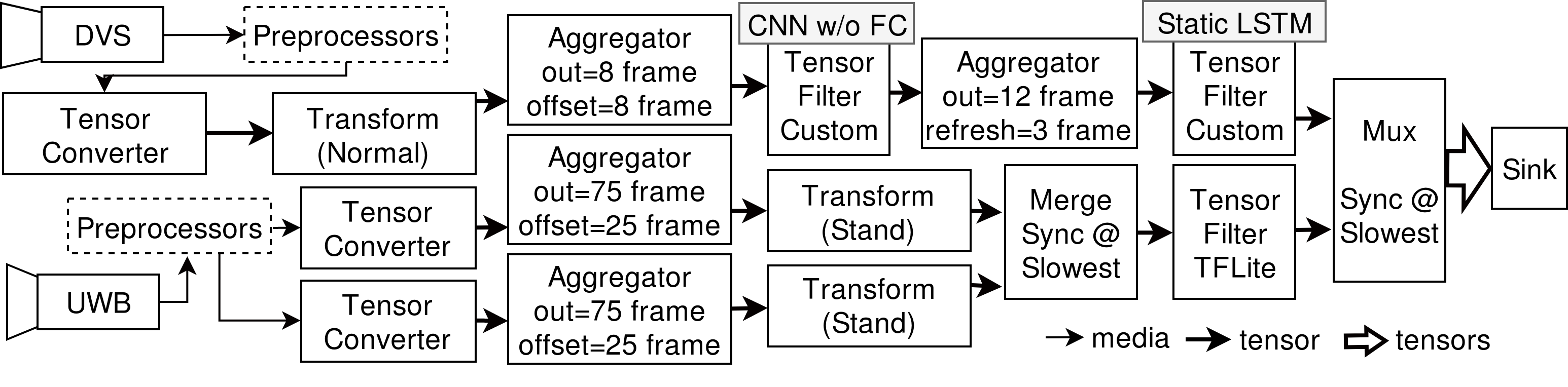}
\caption{ARS Pipeline. The top half, \texttt{B} of Figure~\ref{FIG_EVAL_Activity_Detection}, processes DVS. The bottom half, \texttt{C} of FIgure~\ref{FIG_EVAL_Activity_Detection}, processes UWB.}
\label{FIG_EVAL_LSTM_Sensors}
\end{figure*}

\begin{table*}
\begin{center}
    \centering
    \begin{tabular}{l l c c c c c c r}
     {} & {} & \multicolumn{2}{c}{\texttt{A}) DVS Arg Max} & \multicolumn{2}{c}{\texttt{B}) DVS LSTM} & \multicolumn{2}{c}{\texttt{C}) UWB} & Improvement \\
     Row & Metric & Control & \textit{Nns} & Control & \textit{Nns} & Control & \textit{Nns} & by \textit{Nns} (\%) \\
         &&&&&&&\vspace{-1em}\\
         \hline
         &&&&&&&\vspace{-1em}\\
    1&  LOC & 364 & 169 & 359 & 186 & 383 & 1 & - \\
%    \# mmap & 274 & 97 & 387 & 98 & 375 & 81 & - \\
    2& Total mmap amount (MiB) & 145 & 80 & 161 & 96 & 142 & 58 & 47.5\\
    3& Max \# threads & 12 & 9 & 17 & 9 & 16 & 8 & - \\
%   Time (sec) & 23.694 & 18.321 & 16.491 & 13.094 & 3.548 & 1.293\\
         &&&&&&&\vspace{-1em}\\
    4& CPU (\%), $\infty$ input rate &  31.18 & 28.50 & 56.52 & 49.50 & 44.00 & 46.50 & 5.4\\
    5& Output FPS, $\infty$ input rate & 46.0 & 59.4 & 2.5 & 3.2 & 9.3 & 25.5 & 65.5\\
    6& Output FPS / CPU (\%), $\infty$ input rate & 1.48 & 2.08 & 0.044 & 0.065 & 0.211 & 0.548 & 75.0\\
         &&&&&&&\vspace{-1em}\\
    7& CPU (\%), 30 FPS input rate &  {29.38}  &  {17.35} &  {38.95} &  {28.50} &  {22.10} &  {5.50} & 43.5\\
         &&&&&&&\vspace{-1em}\\
         \hline
    \end{tabular}
    \end{center}
    \caption{Experimental results of ARS with 1100 input frames. FPS denotes frames per second. Nns denotes \textit{nnstreamer}.}
    \label{TBL_RESULT_ACTIVITY}
\end{table*}

ARS is a product experimentally deployed to households and public facilities and will be commercially available soon.
We have applied \textit{nnstreamer} to ARS to simplify the implementation, to improve the performance, and to add more sensors and features.
Because the hardware, \texttt{A} in Table \ref{TBL_SPEC_BOARDS}, has limited resources, processing multiple neural networks and streams is challenging.
To evaluate the performance with the exactly same data, we use preprocessed input data stored as a file.

Figure~\ref{FIG_EVAL_Activity_Detection} describes algorithms for ARS and Figure~\ref{FIG_EVAL_LSTM_Sensors} shows the \textit{nnstreamer} pipeline for ARS.
\texttt{A} in Figure~\ref{FIG_EVAL_Activity_Detection} is the initial CNN-based DVS processing algorithm~\cite{2017NIPS_ActivityDetect},
where each instance of CNN accepts 8 consecutive images with offsets of 4 frames.
Then, it detects activity events by applying the argument max operator to 6 latest CNN results.
\texttt{B} improves \texttt{A} by adding another neural network, LSTM, instead of the simple argument max.
Both \texttt{A} and \texttt{B} use NEON SIMD to accelerate for both \textit{nnstreamer} and conventional implementations.
They are attached as custom sub-plugins of tensor\_filter.
\texttt{C} is a CNN-based neural network implemented with TensorFlow-lite
processing 75 consecutive frames from two standardized streams from UWB, which outputs two streams.

Table~\ref{TBL_RESULT_ACTIVITY} compares the \textit{nnstreamer} implementation and Control, which implements ARS with 
Python and NumPy computation library.
The last column shows the performance improvement achieved by \textit{nnstreamer} with geometric means; i.e., how much CPU time or memory is saved by \textit{nnstreamer} or how much faster is \textit{nnstreamer} than Control.

Row 1 shows the total lines of codes except for neural network models, which indirectly suggests that \textit{nnstreamer} may lower the developmental costs.
Both Control and \textit{nnstreamer} share C binaries of neural network models, \texttt{A} and \texttt{B}.
\textit{Nnstreamer} C codes, \texttt{A} (169) and \texttt{B} (186), are trivial wrappers for the models based on the \textit{nnstreamer} template.
Control Python codes, \texttt{A} (364) and \texttt{B} (359), handle pipelines and threads, which are replaced by a single-line shell script with \textit{nnstreamer} described in the last paragraph of this chapter.
Surprisingly, \textit{nnstreamer} has taken few days of a single developer, with no known bugs, to implement what
Control has taken a few weeks of developers along with known bugs.

\footnotetext{Read / write throughput in GiB/s. Benchmarked with ``pmbw'', \url{https://github.com/bingmann/pmbw}. Big cores are used for \texttt{B} and \texttt{C}.}

Other rows in Table~\ref{TBL_RESULT_ACTIVITY} evaluate the performance.
Row 2 suggests that \textit{nnstreamer} incurs less memory copies, which is critical for embedded devices with low memory throughput.

Row 3 suggests that both implementations try to exploit parallelism with multi-threading.
Row 4, 5, and 6 show the CPU usage (resource consumption), output frames-per-second (throughput), and the efficiency (throughput / resource consumption) respectively in case we have unlimited input rate; i.e., push as many input frames as stream filters process.
They show that \textit{nnstreamer} implementation is far more efficient (75.0 \%) and has higher throughput (65.5 \%) although \textit{nnstreamer} has incurred significantly less time and effort of developers.
Note that the output rate may be slower than the input rate because Aggregator aggregates multiple frames.

The efficiency of real product is shown with Row 7; \textit{nnstreamer} saves 43.5 \% of CPU workloads in average.
Note that \textit{nnstreamer} takes only a quarter of CPU time compared to Control to process UWB.
With higher efficiency, \textit{nnstreamer} enables inexpensive devices to process more sensors and neural networks with reduced developmental costs.

\begin{figure}[t]
\begin{center}
\includegraphics[width=0.99\columnwidth]{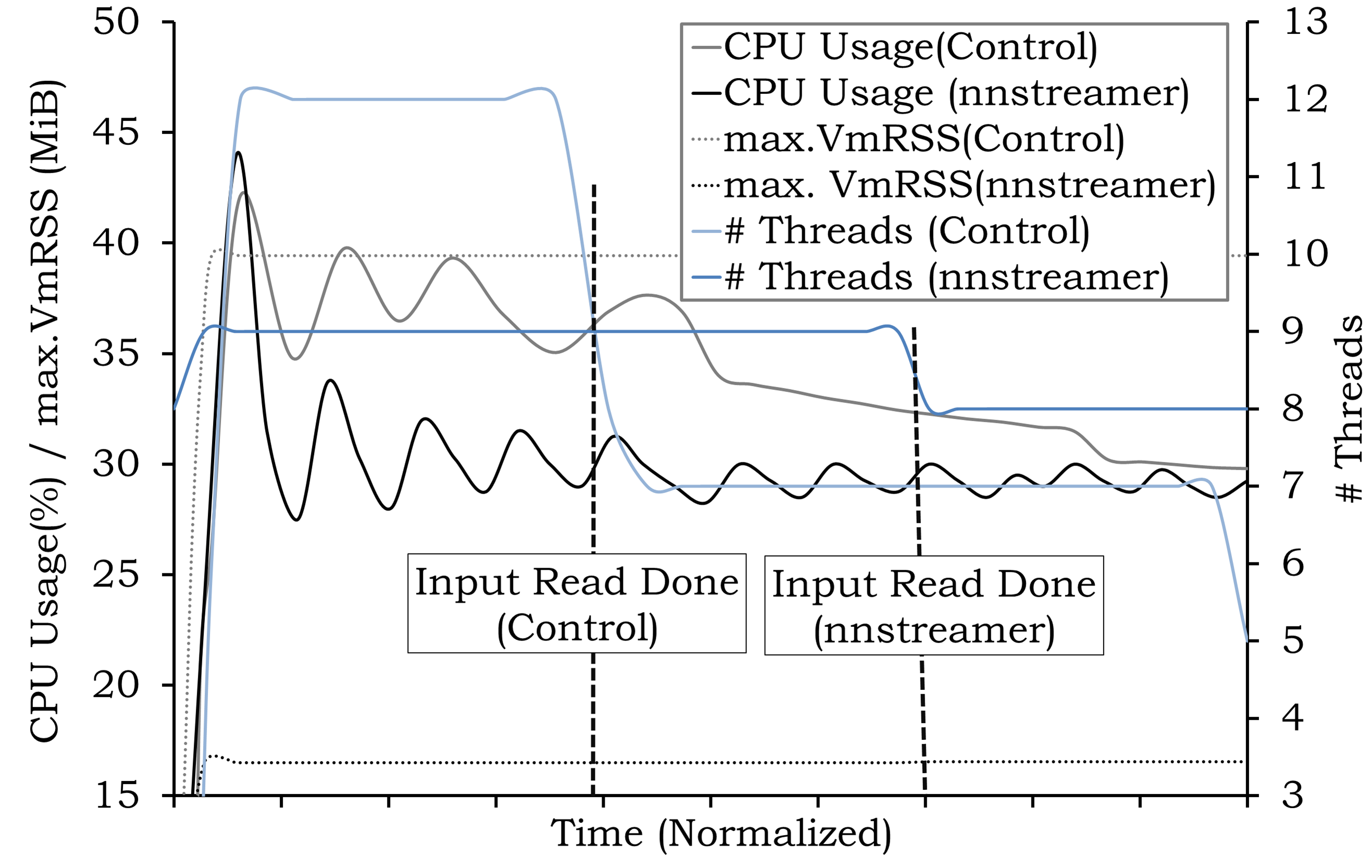}
\end{center}
\caption{Resource utilization over time}
\label{FIG_DVS_ANALYSIS}
\end{figure}

% Analysis 1. Memory

\textit{Nnstreamer} does not incur memory-copy for inter-filter data transmissions.
Even with Tee, there is no memory-copy between a source and sinks of the Tee unless there is an in-place operation--writing the output directly to the input buffer--in a sink.
Therefore, as long as there are no rogue \textit{nnstreamer} custom filters, memory-copy operations are expected to be minimized while developing such optimized but complex pipelines with the conventional method is not so trivial.

% Analysis 2. The efficiency and performance
\textit{Nnstreamer} promotes parallelism with minimal developmental efforts.
The stream pipeline paradigm allows to distribute workloads to multiple elements even if there is a single stream pipeline.
Functional parallelism is promoted by processing elements or sub-pipelines in different processors with less efforts;
the short code in the last paragraph of this chapter is sufficient.
Besides, efficiency is further promoted by regulating processing rates; e.g., a producer will not process faster than if its only consumer.

% Analysys 2-1.
% Additional analysis after the FIG_DVS_ANALYSIS
% A. Less memory footprint (good for embedded!)
% B. Better workload distribution of pipeline. (or temporal distribution)
% B-2. This suits processing continuous streams!!!
%%%%%%%%%% In the figure, (DVS_ANALYSIS), correct # Thread --> # Threads : Done
Figure~\ref{FIG_DVS_ANALYSIS} shows resource utilization over time with unlimited input rates.
The time on x-axis is normalized so that both cases are completed at the end of x-axis.
Both CPU Usage and \# Threads suggest that \textit{nnstreamer} not only incurs less workloads, but also distributes workloads more evenly over time.
Memory, an expensive resource for embedded devices, is significantly saved by \textit{nnstreamer} as well: 39 vs 17 MiB.
%%%%%%%%%%%%%%% @TODO @jijoong-mmon: which case (A/B/C?) is applied for Figure 11? It is case A (DVS+arg. max)

% Analysis 3. Low cost of development (easier implementation)
The major barrier to using complex pipelines with multiple sensors and neural networks has been the difficulties of implementing the pipeline topology with data protections, processing rate regulations, and efficient data transfers.
After the demonstration of the \textit{nnstreamer} pipeline (Figure~\ref{FIG_EVAL_LSTM_Sensors}), ARS developers have abandoned their implementations and started using \textit{nnstreamer} for its simplicity and efficiency.
The following code is a shell script implementing the whole pipeline, which enables to test and modify it quickly along with off-the-shelf GStreamer debugging and profiling tools.

\begin{verbatim}
$ gst-launch-1.0 tensor_mux name=mux ! fakesink  
! tensor_converter ! tensor_trans mode=arith 
! tensor_aggregator in=1 out=8 flush=8 
! tensor_filter frame=custom m=./cnn.so
! tensor_aggregator in=1 out=12 flush=3
! tensor_filter frame=custom m=./lstm.so
! mux.sink_0
tensor_merge name=merge sync-mode=slowest
! tensor_filter frame=tflite m=./uwb.tflite
! mux.sink_1
multifilesrc location="./input_uwb0_%04d.data"
! tensor_converter dim=1:1:32:1 type=float32
! tensor_aggregator in=1 out=75 flush=25
! tensor_trans mode=stand ! merge.sink_0
multifilesrc location="./uwb1_%04d.data"
! tensor_converter dim=1:1:32:1 type=float32
! tensor_aggregator in=1 out=75 flush=25
! tensor_transform mode=stand ! merge.sink_1
\end{verbatim}

\subsection{Experiments with MTCNN}

\begin{figure*}[ht]
\centering\includegraphics[width=0.99\textwidth]{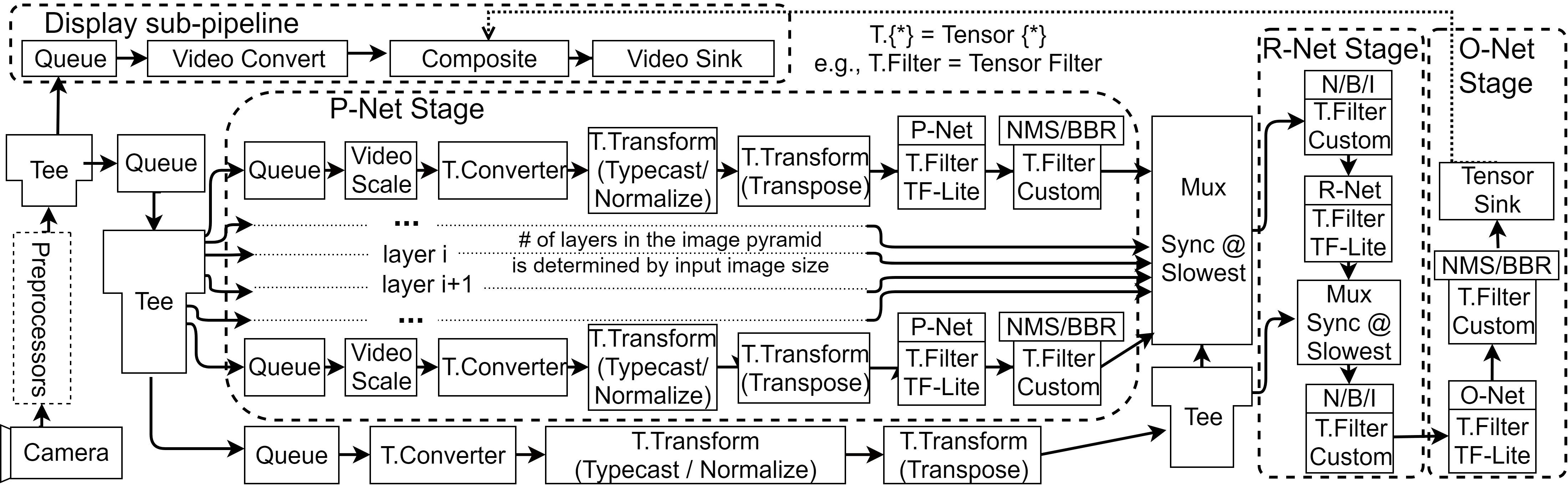}
\caption{Pipeline Topology of \textit{nnstreamer} MTCNN. N/B/I denotes NMS/BBR/Image Patch Generation}
\label{FIG_OVERVIEW_MTCNN}
\end{figure*}

We evaluate \textit{nnstreamer} with MTCNN~\cite{mtcnn-dai2016instance,mtcnn-zhang2016joint} because MTCNN has a complex pipeline topology and real life use cases for many applications; it detects faces and their alignments in a video frame.
We compare the \textit{nnstreamer} and ROS~\cite{quigley2009ros} implementations with Full-HD videos in various devices, \texttt{B} (low/mid-end embedded), \texttt{C} (high-end embedded), and \texttt{D} (PC) described in Table~\ref{TBL_SPEC_BOARDS}.

Figure~\ref{FIG_OVERVIEW_MTCNN} shows the \textit{nnstreamer} pipeline topology of MTCNN, which has two large sub-pipelines.
One, the top dotted rectangle in Figure~\ref{FIG_OVERVIEW_MTCNN}, composites (GStreamer/cairooverlay) and displays (GStreamer/ximagesink) the live camera feed and the result of the other sub-pipeline.
The other sub-pipeline, representing the MTCNN algorithm, is divided into three stages:
Proposal Network (P-Net), Refine Network (R-Net), and Output Network (O-Net).
MTCNN uses a cascade of convolutional networks in an image pyramid.
In other words, each input frame is scaled down to various sizes comprising an image pyramid in P-Net, where each size is represented by a layer.
Therefore, P-Net Stage has multiple streams (one for each layer) and
an instance of P-Net including pre- and post-processing is processed in each stream.
Then, in the order of P-Net, R-Net, and O-Net, each stage is linked and processed, which is 
post-processed by Non-Maximum Suppression (NMS) and Bounding-Box Regression (BBR).

As shown in Figure~\ref{FIG_OVERVIEW_MTCNN}, MTCNN requires image processors: scale, transpose, color space conversion, and data type conversion.
Before applying \textit{nnstreamer}, ROS developers have relied on OpenCV for image processing, requiring additional time and effort.
Besides, the performance enhancement with parallelism has not been successful with the ROS implementation.
To promote parallelism, first, they have tried to adopt the pipe and filter style~\cite{MSPipeFilter} (implementing the stream pipeline paradigm) by dividing the stages as ROS nodes.
However, adding more ROS nodes have increased the message overheads (we use Full-HD videos!) and deteriorated the throughput significantly.
As a result, they have abandoned the pipe and filter style despite of the observation that we can significantly accelerate P-Net by exploiting parallelism.
Applying parallelization libraries such as POSIX Threads to such complex architecture is not a trivial task for many AI project developers, either.

%%%%%%%%%%%%%%%%%%%%%%%%%%%%%%%%%%%%%%%%%%%%%%%%%% NOTE!
% don't use sub-pipeline except for the top (compositor) and the bottom (MTCNN itself). It will confuse readers.

As in ARS, exploiting parallelism has been highly simplified and trivial with \textit{nnstreamer}.
The following code implements a single layer of P-Net\footnote{Unlike ARS, the MTCNN implementation uses C APIs instead of shell scripts in order to construct a pipeline dynamically.}, where \texttt{gst\_parse\_launch}~\cite{GstParseLaunch} constructs a pipeline, which can be a part of a larger pipeline.

\begin{verbatim}
gchar *desc1 = g_strdup_printf (
  "queue ! videoscale ! "
  "video/x-raw,width=%d,height=%d ! "
  "tensor_converter ! "
  "tensor_transform mode=arithmetic " 
    "option=typecast:float32,add:-127.5,"
    "mul:0078125 ! "
  "tensor_transform mode=transpose"
    "option=0:2:1:3 ! " 
  "tensor_filter framework=tflite "
    "model=./pnet.tflite ! " 
  "tensor_filter framework=custom "
    "model=./pnet_pb.so",
  width1, height1);

GstElement *layer1 = gst_parse_launch (
  desc1, NULL);
\end{verbatim}

\begin{table*}
\begin{center}
    \centering
    \begin{tabular}{l l c c}
     {Row} & {Metric} & Control & \textit{nnstreamer} \\
         &&&\vspace{-1em}\\
         \hline
         &&&\vspace{-0.9em}\\
     1 & LOC & 1644 & 1959\\
     2 & Decomposed LOC & {Display: 70, MTCNN: 1574} & {Display: 51, Pipeline: 904, Custom Filters: 1004}\\
     &&&\vspace{-1em}\\
     3 & \# effective threads & 3 & 19\\
     4 & Decomposed \# eff. threads & Input: 1, Display: 1, MTCNN: 1 & P-Net: 14, Others: 5 \\
    \hline
    \end{tabular}
    \end{center}
    \caption{Lines of codes and the number of effective threads except ROS core threads with Full-HD video inputs.}
    \label{TBL_COMP_MTCNN_LOC_THREAD}
\end{table*}

%%%%%%%%%%%%%% @TODO Question to @wook-song Is the "latency" from "stabilized stream"? or "1st frame"?
\begin{table*}[t]
\begin{center}
\centering
\begin{tabular}{l l l c c c c c c c}
     {} & {} & {} & 
     \multicolumn{2}{c}{\texttt{B} / Odroid-XU4} & \multicolumn{2}{c}{\texttt{C} / 8890} & \multicolumn{2}{c}{\texttt{D} / PC} &
     Improvement\\
     {Row} & {Metric} & {Input Rate} &
     Control & \textit{Nns} &
     Control & \textit{Nns} &
     Control & \textit{Nns} &
     by \textit{Nns} (\%) \\
         &&&&&&&&\vspace{-1em}\\
         \hline
         &&&&&&&&\vspace{-1em}\\ 
    1 & End-to-end latency (ms) & 1 FPS & 981.78 & 811.00 & 704.49 & 539.64 & 94.28 & 85.91 & 15.35 \\
    2 & Output rate (FPS) & 30 FPS & 1.01 & 1.73 & 1.48 & 4.02 & 10.41 & 13.76 & 83.21 \\
    3 & Avg. CPU usage (\%) & 30 FPS & 31.85 & 84.60 & 31.15 & 87.10 & 12.90 & 48.74 & - \\
    4 & Memory usage (MiB) & 30 FPS & 136 & 307 & 129 & 474 & 248 & 440 & - \\
    \hline
\end{tabular}
\end{center}
\caption{MTCNN performance with 1000 input frames. FPS denotes frames per second. Nns denotes \textit{nnstreamer}.
}
\label{TBL_RESULT_MTCNN_PERF}
\end{table*}

The above code of a single P-Net layer processes a given size in the image pyramid (\texttt{width1} $\times$ \texttt{height1}), and is linked to Tee and Mux as shown in Figure~\ref{FIG_OVERVIEW_MTCNN}.
Then, each layer is executed as a thread, enabling functional parallelism for P-Net Stage, synchronized by the Mux between P-Net Stage and R-Net Stage. 
The Queue between the two Tees before P-Net Stage allows to drop frames for MTCNN if MTCNN cannot follow the frame rates of incoming video while not affecting the live feed displayed to Video Sink.
For proper and optimized behaviors, we also need to configure queueing policies appropriately; i.e., how buffers are leaked and how many buffers may wait in a queue.

Table~\ref{TBL_COMP_MTCNN_LOC_THREAD} shows the total lines of codes and the number of effective threads and their decompositions.
Control C++ codes has three ROS nodes: Input, MTCNN (70 LOC), and Display (1574 LOC).
The ROS-MTCNN node is a single threaded implementation that does not require any inter-node data transmissions.
We do not count LOC of Input because we reuse open source code for Input.
We do not count threads of ROS core and middleware as they are not effective for parallel executions of MTCNN.

\textit{Nnstreamer} C/C++ codes have a pipeline (Display (51 LOC) and Pipeline (904 LOC) sub-pipelines) and tensor\_filter custom sub-plugins (1004 LOC).
Displays of both implementation have the same functionality and MTCNN of ROS is equivalent to Pipeline and Custom Filters of \textit{nnstreamer}.
Pipeline LOC counts the stream pipeline description of the MTCNN sub-pipeline.
As with ARS, the two custom sub-plugins are the trivial wrappers based on the \textit{nnstreamer} template.
Compared to Control (1644), \textit{nnstreamer} (1959) has more LOC.
However, considering that only a few more trivial codes are required (333 LOC) to achieve parallelism in such a complex pipeline, the result suggests that \textit{nnstreamer} allows to make development cost lower with higher efficiency and better quality than the conventional method. 
Besides, compared to Control, \textit{nnstreamer} developers have added a lot more features including extensive error handling, appropriate frame dropping, dynamic number of layers, and dynamic input video sizes and rates along with the reliability to handle camera inputs continuously.

Control consists of 3 single-threaded ROS nodes, incurring 3 effective threads.
Besides, most computation happens with MTCNN node, which implies that there is little effective parallelism with Control except for OpenCV functions.
\textit{Nnstreamer} has much higher number of effective threads (19) although we do not write any multi-threaded codes explicitly.
Particularly, P-Net Stage has 14 threads automatically and dynamically (according to the size of input images) created by the framework, which implies that \textit{nnstreamer} highly promotes parallelism easily and efficiently.

\begin{figure}[t]
\begin{center}
\includegraphics[width=0.99\columnwidth]{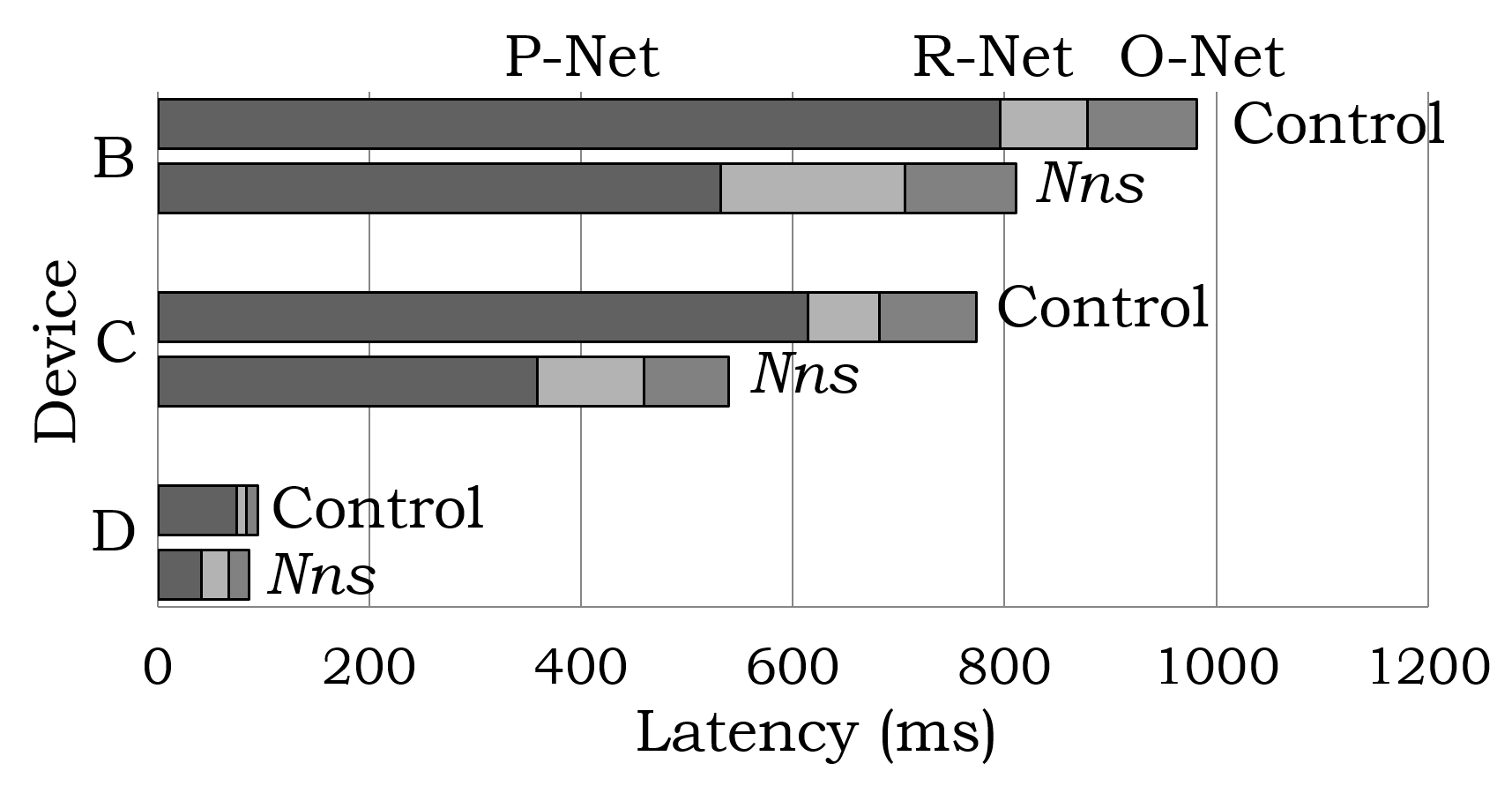}
\end{center}
\caption{Breakdown of the face detection latency.}
\label{FIG_MTCNN_LATENCY_BD}
\end{figure}

Table~\ref{TBL_RESULT_MTCNN_PERF} compares the performance of the two implementations in various devices.
Row 1 shows the end-to-end latency between the sensor input to the display output for a single image frame with the input rate (1 FPS) slow enough to have a single frame in a pipeline.
Moreover, Figure~\ref{FIG_MTCNN_LATENCY_BD} shows the decompositions of latency, which suggests that most of the latency reduction comes from P-Net.
More specifically, the performance, defined as $1 / \textrm{latency}$, of P-Net Stage is
improved by 49.73\%, 71.53\%, and 80.83\% while the performance of other stages is
degraded by 33.44\%, 12.54\% and 55.49\% for Device \texttt{B}, \texttt{C}, and \texttt{D}, respectively. 
Even in such a case, \textit{nnstreamer} improves the performance by 20.15 \% (up to 30.55 \% for Device \texttt{B}) because P-Net Stage is the most critical part.
Note that P-Net Stage of ROS-MTCNN contributes 82.29\% to the total latency.
In other words, Row 1 shows the performance comparison without the impact of pipelined data-parallelism~\cite{1990PipelinedDataPar}, but with the impact functional parallelism of P-Net despite of less efficient R-Net and O-Net.

Row 2 shows the output rates from 30 frames per second input of a Full-HD camera, processing Full-HD images with MTCNN.
This allows pipelined data-parallelism, allowing processing different frames in filters.
As a result, the performance improvement by \textit{nnstreamer} is more significant: 83.21\% (up to 171.62\% for Device \texttt{B}.)
%Therefore, as expected, the performance gain is mostly from the pipeline data-parallelism of P-Net stage: 44.70\%, 41.70\%, and 33.21\% for Device \texttt{B}, \texttt{C}, and \texttt{D}, respectively.

Note that the output frame rate is slower than the input rate; thus, input frames are often dropped in front of P-Net with the \textit{nnstreamer} implementation.
However, the frames for display compositing are not dropped and the final display output keeps showing 30 FPS video.
On the other hand, the ROS implementation has no such luxury and it cannot create 30 FPS display by dropping frames appropriately.

Analysing the effect of functional parallelism, we have mentioned that the performance of R-Net and O-Net have not been improved by the \textit{nnstreamer} implementation; it is rather deteriorated as shown in Figure~\ref{FIG_MTCNN_LATENCY_BD}.
This is mainly because the two custom tensor\_filter sub-plugins (N/B/I) in R-Net Stage incur memory copies of Full-HD frames in order to generate small image patches for R-Net and O-Net.
On the other hand, Control uses a global variable to avoid memory copies for the image patch generation.
As a result, with devices of higher memory bandwidth, the performance deterioration in R-Net is reduced; 16.21 ms for Device \texttt{D}, 33.66 ms for Device \texttt{C}, and 91.98 ms for Device \texttt{B}.

The \textit{nnstreamer} implementation uses more CPU time to process faster.
Because, output rates of both implementations have not reached the input rates, it would be better to consume more CPU time to process more frames per second.
The memory consumption is higher with the \textit{nnstreamer} implementation although it is within the acceptable limit.
However, MTCNN-\textit{nnstreamer} consumes more than 10 times of the memory consumed by ARS-\textit{nnstreamer}, which implies that the pipeline implementation is wasting the memory, probably by queues: we have attached too many queues that may store Full-HD frames.
For the image pyramid of MTCNN, it would be significantly efficient (for both CPU and memory) if we write a custom tensor\_filter sub-plugin that generates multiple layers of images directly from an input stream.

%% file: sections/S06_FutureWork.tex
The following filters and features are in our short-term plan.

\begin{itemize}
    \item \textbf{tensor\_ros\_sink/src} allow to exchange tensor streams with Robot OS (ROS) \cite{quigley2009ros} nodes for ROS1 and ROS2.
    \item \textbf{tensor\_save/load} store tensor streams as files and load the files as tensor streams.
    \item \textbf{tensor\_protobuf\_sink/src} allow to use Google's protobuf as a transport layer.
    \item \textbf{tensor\_source} accepts sensor data not compatible with GStreamer including LIDARs, RADARs, and others.
    \item More NNFWs for \textbf{tensor\_filter}: Caffe/Caffe2 and Keras.
    \item Make \texttt{other/tensor(s)} a GStreamer standard type.
\end{itemize}

There are long-term plans as well.
First, we will apply the actor model~\cite{agha1985actors} to make pipelines fully-distributed.
Robotics projects often incur multiple computers with different roles,
which requires filters of a pipeline distributed across multiple computers.

Another long-term plan is to allow retraining a pipeline directly solely on device.
Federated Learning~\cite{FederatedLearning} allows to update models by retraining models in cloud based on the data gathered from edges; however, it is difficult to apply Federated Learning if protected privacy data is required for training and the data cannot be transmitted.
Therefore, we are sometimes required to retrain neural networks directly in edge devices.

%% file: sections/S07_Conclusions.tex
From the experiences of on-device AI applications, our conjecture is that applying stream processing paradigm can greatly improve the performance, and the implementation productivity.
By implementing and deploying \textit{nnstreamer}, we show that such improvements can be achieved successfully for on-device AI applications: higher throughput and easier implementation with more features and higher code quality.
As a side effect, traditional multimedia developers can now employ arbitrary neural network models in their own multimedia pipelines with \textit{nnstreamer}.

This work is being deployed for commercial products by the affiliation of the authors.
We plan to adopt \textit{nnstreamer} to a wide range of devices by making it the standard intelligence framework of Tizen OS.
This work is compatible and tested with many software platforms including Tizen, Android, and Ubuntu and is open source software licensed in LGPL.